\documentclass[%
 aip,
 amsmath,amssymb,
preprint,%
]{revtex4}

\pdfoutput=1

\usepackage{graphicx}
\usepackage{dcolumn}
\usepackage{bm}

\usepackage{graphicx}
\usepackage{subfig}

\begin{document}


\title[Monolithic semiconductor hemispherical microcavities for efficient single photon extraction]{Monolithic semiconductor hemispherical microcavities for efficient single photon extraction}

\author{G. C. Ballesteros}
\email{gb173@hw.ac.uk}
\author{C. Bonato}%
\email{c.bonato@hw.ac.uk}
\author{B. D. Gerardot}%
\email{b.d.gerardot@hw.ac.uk}
\affiliation{Institute of Photonics and Quantum Sciences (IPaQS), Heriot-Watt University, Edinburgh, EH14 4AS, UK}

\homepage{http://qpl.eps.hw.ac.uk/}

\date{\today}

\begin{abstract}
We present a monolithic semiconductor microcavity design for enhanced light-matter interaction and photon extraction efficiency of an embedded quantum emitter such as a quantum dot or color center. The microcavity is a hemispherical Fabry-Perot design consisting of a planar back mirror and a top curved mirror. Higher order modes are suppressed in the structure by reducing the height of the curved mirror, leading to efficient photon extraction into a fundamental mode with a Gaussian far-field radiation pattern. The cavity finesse can be varied easily by changing the reflectivity of the mirrors and we consider two specific cases: a low-finesse structure for enhanced broad band photon extraction from self-assembled quantum dots and a moderate-finesse cavity for enhanced extraction of single photons from the zero-phonon line of color centers in diamond. We also consider the impact of structural imperfections on the cavity performance. Finally, we present the fabrication and optical characterisation of monolithic GaAs hemispherical microcavities.
\end{abstract}

\maketitle

\section{Introduction}
Several applications of quantum technology rely on single photon emission \cite{somaschi_near-optimal_2016,ding_-demand_2016,dada_indistinguishable_2016}. Deterministic sources of pure and indistinguishable single photons are a crucial prerequisite to linear optics quantum computing \cite{knill_scheme_2001} and restricted models of non-universal quantum computation \cite{he_scalable_2016}. Long-distance quantum communication using quantum repeaters \cite{briegel_quantum_1998} relies on a coherent interface between photons and matter qubits \cite{gao_coherent_2015}, such as spins associated to quantum dots \cite{delteil_generation_2015} or colour centres \cite{hensen_loophole-free_2015}. In both cases, efficient photon extraction from the emitter is crucial, either to reach the fidelity threshold for fault-tolerance or to achieve a significant communication speed.

The efficiency of single photon extraction can be enhanced by a structure that re-distributes the emitter radiation into a highly-directional spatial profile which is easy to collect. For some applications, collection must be enhanced over a broad spectral bandwidth.
A possible approach is to use non-resonant structures such as solid immersion lenses (SIL) or nanowires. A micro-scale solid immersion lens \cite{hadden_strongly_2010,marseglia_nanofabricated_2011,gschrey_highly_2015} significantly reduces total internal reflection by the high-index host material, leading to increased, though far from unity, photon extraction over a very wide spectral region. Broadband extraction with efficiency close to unity can be achieved in a nanowire  structure \cite{munsch_dielectric_2013,bulgarini_nanowire_2014}. This configuration can however be problematic when fabricating electrical contacts or exploiting the spin of the emitter, which is subject to dephasing induced by the nanometric proximity to the nanowire walls.

A different approach involves coupling the emitter to a resonant structure, which exploits the Purcell effect to reduce the radiative lifetime of the source and increase the rate of excitation/emission events. Additionally, it favours coupling to a single mode that can be efficiently collected over guided modes within the plane of the sample. High-Q resonant microcavities, such as micropillar and photonic crystal structures, are particularly valuable when the emitter features strongly phonon-broadenend transitions, e.g. the nitrogen-vacancy colour centre in diamond. In this case, the coherent zero-phonon line can couple efficiently to the resonant mode, achieving a significant Purcell factor, which results in a strongly enhanced emission at the expense of the phonon-broadened transitions \cite{riedel_deterministic_2017}.
High-Q microcavities, however, present several technological challenges, since they require precise spectral tuning of the cavity resonance on the emitter frequency. While fully-tunable open microcavities have been demonstrated \cite{greuter_small_2014,dolan_femtoliter_2010,riedel_deterministic_2017, najer_gated_2018, vogl_space_2019}, tunability comes at the price of reduced mechanical stability and increased sensitivity to vibrations \cite{bogdanovic}

Here, we propose a monolithic hemispherical cavity  which addresses the mechanical stability challenges outlined above. It is a flexible design which provides stable cavity modes with lifetimes that can be tailored by modifying the reflectivity of the enclosing mirrors. For narrowband emitters, such as InAs/GaAs quantum dots, a weakly-resonant structure with only a moderate Purcell factor is sufficient to achieve a very strong spatial directionality of the emitted photons. We discuss three design variations, adapting the quality factor to the properties of the specific emitter under consideration. We also fabricate and optically characterize a monolithic GaAs hemispherical microcavity to experimentally verify optical resonances in the structure.

 \section{Concept and theory}

\begin{figure}[htbp]
\centering
\includegraphics[width=\linewidth]{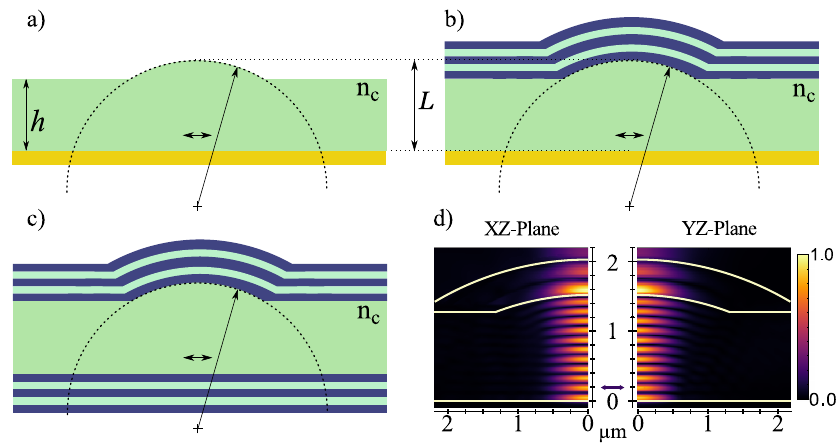}
\caption{Three examples of hemispherical micro-cavity designs. The dipole moment of the emitter is contained in the XZ plane. The cavity with index of refraction $n_c$, has length $L$. The height of the remaining slab is given by $h$. Case a) low-Q case: a hemispherical cap is etched onto a membrane embedding the emitters, coated with a bottom gold mirror. b) medium-Q case: a DBR comprising 5 layers is added on the top. c) high-Q case: including DBRs at the top and the bottom. d) Intensity profile for the fundamental cavity mode, in the medium-Q case (FDTD simulations).}
\label{fig:microsil_design}
\end{figure}

Our basic design consists of a hemisphere with radius of curvature $R$ milled on the surface of a material slab of thickness $L$ with embedded optical emitters (Fig. \ref{fig:microsil_design}). In the lower-Q version (Fig. \ref{fig:microsil_design}a), reflection from the top air-material interface provides the required longitudinal optical confinement. In this case, a gold mirror can be used at the bottom of the structure since losses are dominated by transmission through the upper interface. This design is suitable for narrowband emitters such InAs/GaAs quantum dots or silicon-vacancy centres in diamond and is appealing due to its simplicity and ease of fabrication. In order to reduce losses through slab guided modes, the quality factor can be increased by adding a 2.5-period conformal DBR on the top surface to increase confinement and Purcell enhancement ("medium-Q" configuration). Finesse can be further increased by adding additional DBR layers to the top surface and substituting the bottom gold mirror with a DBR to reduce non-radiative losses ("high-Q" design). The high-Q version is suitable to enhance radiation into the coherent zero-phonon line for solid-state emitters \cite{grange_reducing_2017, riedel_deterministic_2017}.


All three versions are based on a monolithic design that minimizes mechanical instability issues associated with open cavities and is fully compatible with the integration of electrical contacts for charge-state control and electrical tuning of emitter transitions\cite{warburton_optical_2000,ma_efficient_2014}. The structure can be fabricated via Focused Ion Beam (FIB) milling \cite{hadden_strongly_2010,marseglia_nanofabricated_2011} or gray-scale lithography and dry etching \cite{gschrey_highly_2015,choi}. The proposed designs require minimal etch depths, reducing fabrication errors and crystal damage.\\

Our microcavity can be modeled based on standard hemispherical laser cavities \cite{siegman_lasers_1986}, which we briefly summarize here to discuss the trade-offs associated with each parameter. A hemispherical cavity supports stable modes if the center of curvature of the spherical cap lies below the flat mirror, i.e. $R>L$, as shown in Fig.~\ref{fig:microsil_design}. The cavity features Laguerre-Gaussian modes, indexed by their axial ($q$), radial ($n$) and azimuthal ($m$) numbers as TEM$_{q,n,m}$ owing to their transverse electromagnetic nature. The resonance condition is given by:

\begin{equation} 
\frac{\Delta\phi_{\mathrm{rt}}}{2}-(n+m+1)\cos ^{-1} \left ( \sqrt{ 1-\frac{L}{R}}\right)=q \pi
\label{eq:resonance_condition}
\end{equation}
where $\Delta\phi_{\mathrm{rt}}$ is the phase accumulated on a round-trip through the cavity. An emitter close to a field antinode experiences a reduction in radiative lifetime due to the Purcell effect. This increases the source efficiency since photons are preferentially emitted into the cavity modes rather than into modes guided in the sample. The Purcell factor increases with the cavity quality factor as $F_p = \frac{3}{4 \pi^2}\left(\frac{\lambda}{n}\right)^3\left( \frac{Q}{V}\right)$. 
High directionality in optical emission stems from the modes supported by the cavity. Since the far-field divergence angle of a fundamental transverse mode TEM$_{q\, 0\, 0}$ is:
\begin{equation}
\theta_{1/e}=\sqrt{ \frac{\lambda}{\pi\sqrt{L(R-L)}}},
\label{eq:beam_divergence}
\end{equation}
the beam divergence is minimized when $R=2L$.\\
To evaluate the overall performance of photon extraction we define the figure of merit (FOM):
\begin{equation}
FOM=\eta_\mathrm{coll}^{\mathrm{NA}} \times \eta_\mathrm{ext} \times F_p,
\label{eq:fom}
\end{equation}
where $\eta_\mathrm{coll}^{\mathrm{NA}}$ is the fraction of power emitted into the far-field that can be collected with a lens with numerical aperture NA:
\begin{equation}
\eta_\mathrm{coll}^{\mathrm{NA}}=\frac{\int^{\arcsin(\mathrm{NA})}_0 \mathrm{d}\theta \; P(\theta) \sin \theta}{\int^{\pi/2}_0 \mathrm{d}\theta \; P(\theta) \sin \theta}.
\label{eq:coll_eff}
\end{equation}

$\eta_\mathrm{ext}$ is the extraction efficiency computed as the fraction of power emitted that does not couple to guided modes and $F_p$ is the Purcell factor. The figure of merit is proportional to the power emitted by the micro-cavity for a given excitation power, providing a valuable performance estimator. As a reference, the emission  of a dipole embedded in a homogeneous medium collected using index-matched lenses with $NA = 1$ would result in FOM $=0.5$.

The depth of the hemispherical etch, $L - h$ (Fig. 1), determines both the order of the highest transverse mode that can exist within the structure and the fabrication complexity. Shallower structures allow faster fabrication reducing cost and beam drift problems during FIB etching. Higher order modes are spatially larger and therefore require large hemispherical surfaces. The etch depth $L-h$ must be as small as possible to strongly suppress higher order transverse modes and minimize fabrication time, while avoiding the apodization of the fundamental transverse modes.

\section{Design and performance analysis}
For ease of comparison, we consider designs based on a slab of GaAs (refractive index $n_c=3.48$); extension to other materials, such as diamond or silicon carbide, is straightforward. The cavity dimensions are listed in Table \ref{table:dimensions}.

According to Eq.~\ref{eq:beam_divergence}, a low beam divergence is achieved by a long cavity length. On the other hand, a short cavity increases Purcell enhancement. As a trade-off, we choose a 11$^\mathrm{th}$ order cavity, designing $L$ to set the resonance wavelength at $940$ nm (Eq.~\ref{eq:resonance_condition}). Efficient coupling to a resonant mode is achieved when the emitter is located at an antinode of the electric field. In hemispherical cavities, the strongest field antinode is located closest to the flat mirror at a distance of approximately half the resonant cavity wavelength. However, in the following discussion we locate the emitter at the position of the second anti-node to avoid surface-induced decoherence or coupling to surface plasmon polaritons in the case of a metallic mirror.  The depth of the etch, i.e. $L-h$ is no more than 190 nm, which greatly facilitates fabrication.

The electromagnetic modes of the structures in Fig.~\ref{fig:microsil_design}a-c were simulated using a commercial finite-difference time-domain software (Lumerical). Eq.~\ref{eq:resonance_condition} predicts a resonant wavelength of 910 nm for the $TEM_{11,0,0}$ mode. We attribute the small deviation in wavelength to reflections at the interfaces and to the apodization of the mode at the top hemisphere.

\begin{figure}[htbp]
\centering
\includegraphics[width=\linewidth]{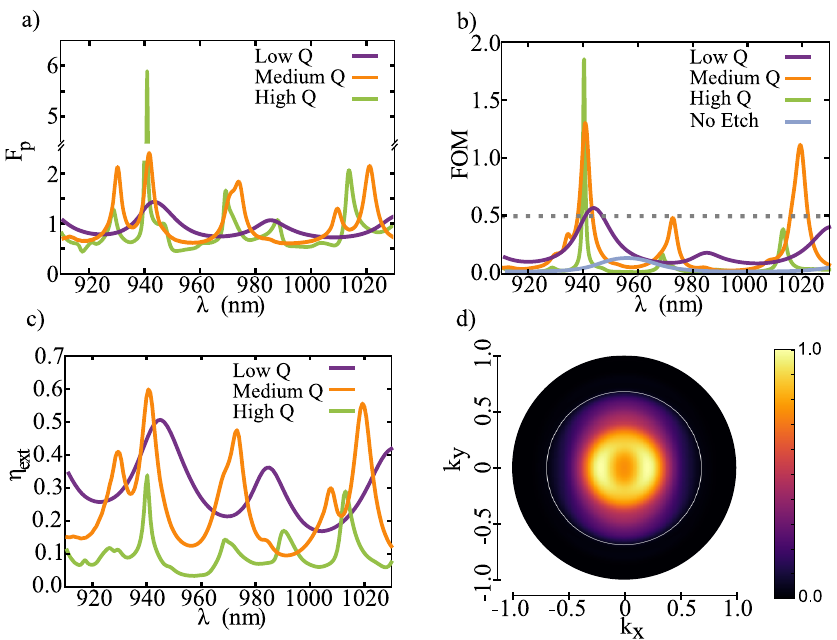}
\caption{(color online). FDTD simulation results. All structures have been adjusted in order to have the $TEM_{11,0,0}$ mode at an excitation wavelength of 940 nm.  (a) Purcell factor. (b) FOM using Eq.~\ref{eq:fom}. The blue line shows the FOM for an un-etched structure. The grey horizontal dotted line represents the figure of merit for a dipole embedded in a homogenous medium with NA=1 and perfect extraction efficiency. (c) Extraction efficiency. (d) k-space representation of the far-field emission for the medium-Q cavity. The inner white circle marks the wavevectors collected by a lens with NA=0.68}
\label{fig:numerical_results}
\end{figure}

The simulated Purcell enhancement (Fig.~\ref{fig:numerical_results}a), exhibits several peaks corresponding to the fundamental transverse modes TEM$_{11,0,0}$ ($\approx 940$ nm), TEM$_{10,0,0}$ ($\approx 1020$ nm) and to higher order transverse modes TEM$_{(10,11), n, m}$. These modes are weakly excited due to the presence of field antinodes along the axis of the cavity and the apodization of the mirror. As expected, higher mirror reflectivity leads to higher Purcell factor and FOM.  Mirror reflectivity also affects the extraction efficiency: as confinement increases, the relative density of photonic states between guided modes and the resonant cavity mode decreases. This leads to a better coupling into the mode of interest and increased extraction efficiency for the low and medium quality factor structures. The high quality factor structure has its extraction efficiency limited by light lost at large angles on the DBR. This effect is evidenced by the highly-directional far-field emission in  Fig.~\ref{fig:numerical_results}d for the TEM$_{11,0,0}$ resonance medium-Q case. The emitted Gaussian profile with an angular spread of $\pm 25^\circ$ (corresponding to NA = $0.42$), leads to a a 20-fold enhancement of the figure of merit of the device when compared to the unetched case.   
On the other hand, increased mirror reflectivity decreases the operation bandwidth $\Delta \lambda$: in our simulation $\Delta \lambda$ decreases from $14$ nm to less than $1$ nm from the low-Q to the high-Q case. 

\begin{figure}[htbp]
\centering
\includegraphics[width=1.0\linewidth]{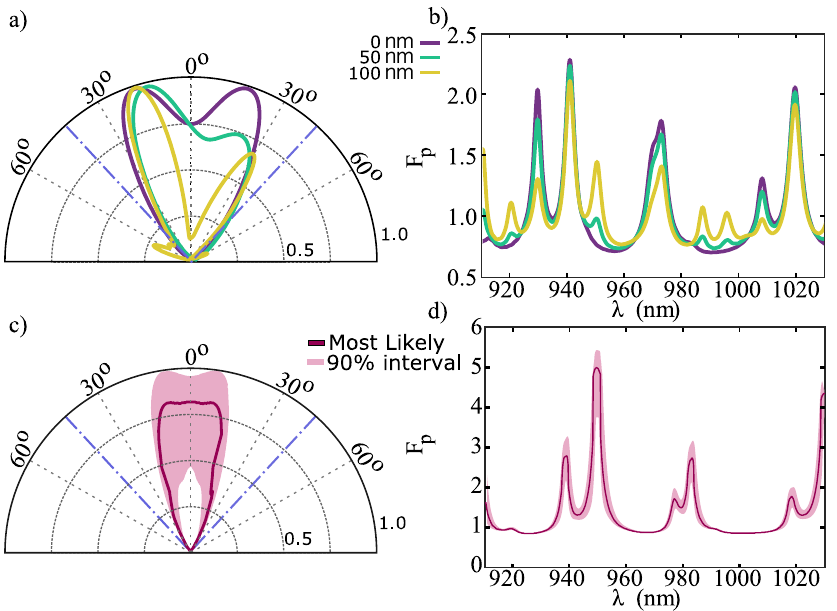}
\caption{Robustness against source displacement and surface roughness. a) Emission profile and b) Purcell factor for a source displaced from the center of the cavity by  0, 50 and 100 nm. Only a slight degradation is observed. c) and d) Monte Carlo study of the effect of surface roughness on the performance of the TM micro-cavity. The shaded are represents the 90\% interval. The continuous line represents the most probable value and the dashed line the performance of the smooth device.}
\label{fig:imperfections}
\end{figure}

The robustness of this design to source misplacement is an important technological consideration. As evidenced by the field profiles in Fig.~\ref{fig:microsil_design}~d, the beam waist is located at the top of the gold reflector with a mode radius of 238 nm. Any emitter within approximately half this distance from the center of the cavity couples sufficiently to the cavity mode. Figure~\ref{fig:imperfections} reports simulation results in the medium-Q case for a dipole source shifted from the cavity center by up to 100 nm. As expected, the pointing angle of the radiation pattern deviates from the optical axis.  However, the resulting Purcell factor does not degrade significantly. The positioning imperfection investigated by our simulations is larger than what has been experimentally achieved with deterministic fabrication processes \cite{sapienza_nanoscale_2015}.
A second set of simulations addresses the effect of surface roughness on the device performance.  A Monte Carlo analysis, comprising  2500 simulations for a root mean square roughness of 5 nm in the case of a medium Q cavity, is reported on the bottom of Fig. ~\ref{fig:imperfections}. To limit the computing time,  2D FDTD simulations were used since such an analysis with 3D simulations is not practical. Comparing the results of the two methods, the significant differences in Purcell factor are due to the cylindrical ($\sim 1/r$ decay) nature of waves in 2D simulations versus spherical waves ($\sim 1/r^2$ decay) in 3D simulations.  Figure~\ref{fig:imperfections} shows that even for a high value of the roughness the device is expected to operate close to optimum performance with a high probability.

\section{Fabrication and Characterization}

Samples based on the low-Q design were fabricated and optically characterized. Unlike the design presented above the fabricated sample made use of a DBR as bottom reflector instead of a layer of Au. Due to the low quality factor of the device, modifying the bottom mirror only required a small re-optimization of the height of the cavity.

\begin{figure}
\centering
\includegraphics[width=0.8\textwidth]{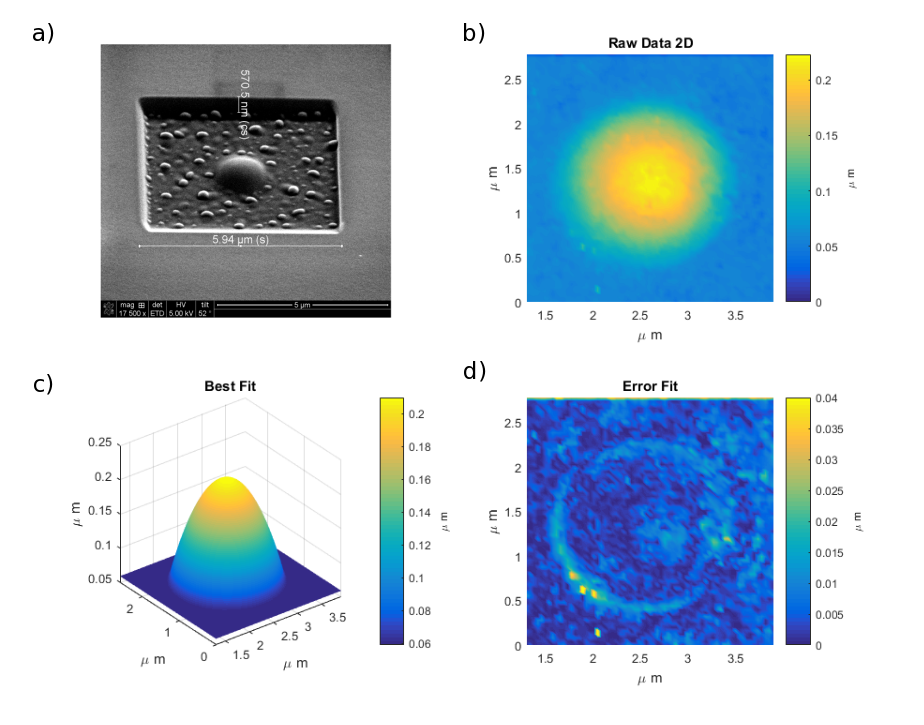}
\caption{(a) Close-up scanning electron microscope image of $\mu$SILs. The small droplets next to the the structure are caused by Ga redeposition. 
(b) Raw AFM data. (c) Fit to AFM data . (d) 
Difference between the best fit surface and the AFM data. The structure shown here
has a radius of curvature of  of 3.2~$\mu$m and the depth  of the etch is a 151 nm. The root mean square value of the roughness of the $\mu$SIL based on the fit and the AFM values is of  4~nm which is much smaller than the operation wavelength.
}
\label{fig:closeupmicrosil}
\end{figure}

The samples were fabricated using water assisted FIB. The FIB current was
 50~pA in each case, providing a good compromise between milling speed and
 Ga redeposition. The depth of the etch was designed to be 180~nm. To facilitate
 calibration of the fabrication parameters, several other structures with etch
 depths of 150~nm, 160~nm, 170~nm were made. The milling time for each
 structure was 20 minutes approximately; the only differences were the number
 of passes taken by the FIB mill. Figure~\ref{fig:closeupmicrosil} show an SEM
 image of one of the devices. One can observe small droplets inside the milling
 area due to gallium redoposition. The milled structures were then characterized
 via AFM. The measured profiles were fitted to a hemisphere on a plane. The
 height ($h$) as a function of the $(x,y)$ coordinates is given by:

\[
 h(x,y) =
  \begin{cases} 
      \hfill z_s + \sqrt{R^2 - r^2 }    \hfill & \text{ if $ (z_s + \sqrt{R^2 - r^2 })>z_p$} \\
      \hfill z_p \hfill & \text{ if $ (z_s + \sqrt{R^2 - r^2 }) \le z_p$ } \\
  \end{cases},
\label{eq:fit}
\],

where $r$ is the radial coordinate, $z_s$ is the position  of the sphere center with respect to the
coordinate origin, $z_p$ is the $z$ position  of the flat area and $R$ is the
radius of curvature of the spherical section. An example of such fit can be seen 
in Fig.~\ref{fig:closeupmicrosil}. The small difference between the fit function
and the measured AFM profile demonstrates how well the required structures
can be fabricated.

\begin{figure}
  \centering
  \includegraphics[width=0.8\textwidth]{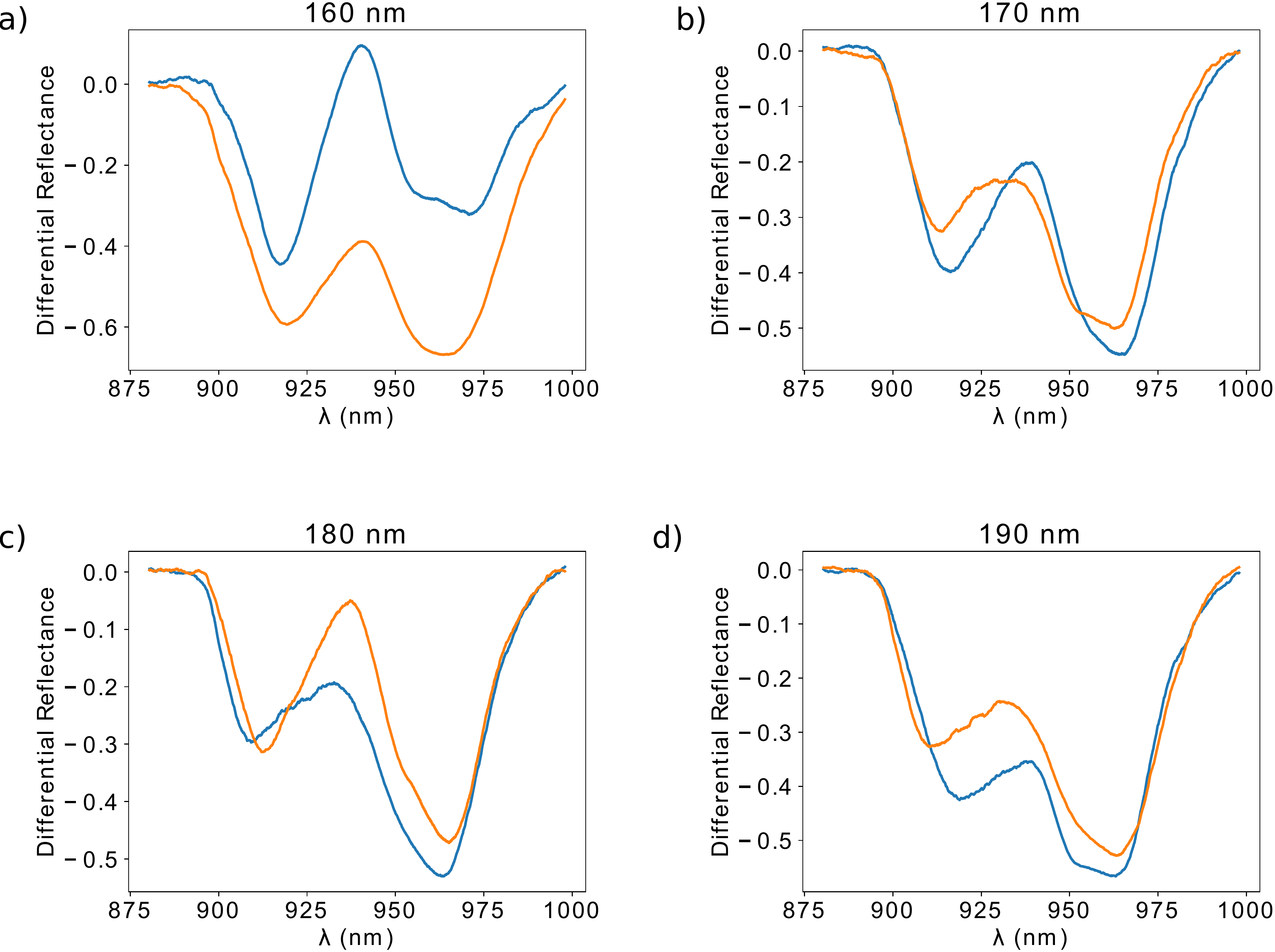}

 \caption{Room temperature differential reflectance spectroscopy on microSILs.
 The target depth etch on the FIB was 160~nm (a), 170~nm (b),
 : 180~nm (c) and 190~nm (d). Two structure were fabricated for each target
 depth. Orange and blue curves show the optical characterization for each pair.}

\label{fig:diff_reflectance_usil}
\end{figure}

We optically characterized the cavity modes by measuring the 
reflectance of the structure using a confocal microscope setup. The samples were illuminated with a fibre coupled infrared LED lightsource with emission centered at 940 nm.
Figure~\ref{fig:diff_reflectance_usil} shows the results of the
differential reflectance measurements taken on eight of the hemispherical
microcavities. The differential reflectance is defined as:

\begin{equation}
    \frac{I_\mathrm{SIL}(\lambda) - I_\mathrm{ref}(\lambda)}
         {I_\mathrm{ref}(\lambda)},
\end{equation}
where $I_\mathrm{ref}(\lambda)$ is a measurement of the reflected spectra made
on a flat area adjacent to the structure and $I_\mathrm{SIL}(\lambda)$ is
a measurement of the reflected spectra when focusing at the center of the
structure. Each measurement shows two clear dips at around 920~nm and 960~nm. These
correspond to the two broad resonances of the low-Q design (purple curves in
Fig.~\ref{fig:numerical_results}). A shift of 20~nm with respect to the
simulations can be observed which could be due to fabrication inaccuracies. The
FWHM of the measured resonances is of 24~nm and 31~nm.

\section{Conclusion and Outlook}
These results demonstrate the potential of the micro-cavity design for a large range of applications. In the case of narrowband emitters, the low and medium Q configurations are preferable due to their ease of fabrication, broadband operation and resilience to imperfections such as scattering losses. Additionally, a low-Q design is more favorable for the implementation of spin-photon interfacing protocols based on spin-selective circularly-polarized optical transitions \cite{bonato_cnot_2010}, as for InAs/GaAs quantum dots. Typically, strain and fabrication imperfections break the degeneracy of the fundamental mode into a pair of linearly-polarized modes. In the case of narrow resonances, when the frequency splitting between the two linearly-polarized modes is larger than the linewidth, this system cannot support the required circularly-polarized modes \cite{bonato_tuning_2009}. On the other hand, the wider bandwidth associated with the low/medium-Q configurations significantly reduces this problem.

The high-Q configuration is particularly suitable for emitters with incoherent broadband emission accompanying the coherent zero-phonon line, such as nitrogen-vacancy centres in diamond. Such emitters predominantly radiate into incoherent phonon sidebands, which hardly exhibit any Purcell enhancement when coupled to a cavity. Phonon-broadened emission cannot be used in quantum interference experiments, providing a limit to the success rate for measurement-based spin entanglement protocols \cite{bernien_heralded_2013}. In this case, a high-Q microcavity approach as shown in Fig. \ref{fig:microsil_design}c is desirable since it can drastically enhance the coherent zero-phonon emission at the expense of the phonon-broadened transitions \cite{grange_reducing_2017, riedel_deterministic_2017}.


 In the high-Q scenario, an additional challenge is the requirement for spectral tunability, due to the narrowband operation range. Compared to the open microcavity case, in our design transverse optical confinement is provided by the hemispherical surface milled on top of the emitter. Consequently, no in-plane scanning capability is required and the only parameter that needs to be tuned is the cavity resonance frequency. This can be achieved by a single piezo-element that controls the distance between the sample and a (detached) bottom mirror. Requiring only one movable element, this configuration can be expected to be mechanically more stable than an open cavity featuring motion along three axes with a stack of piezoelectric elements. 

In summary, we have presented a novel microcavity design suitable for efficient photon extraction from optical emitters embedded in high index of refraction mediums such as III-V quantum dots and luminescent point defects in wide-bandgap semiconductors.  One example of the proposed structure has been fabricated in a GaAs sample and optical and structural characterization reveals good agreement with the targeted design.

\section{Funding Information}

This work has been supported by the Engineering and Physical Sciences Research Council (EPSRC) under the grants: EP/I023186/1, EP/L015110/1, EP/S000550/1, EP/P029892/1 and by the European Research Counil (ERC) under grant number 725920. B.D.G. thanks the Royal Society for a Wolfson Merit Award and the Royal Academy of Engineering for a Chair in Emergent Technologies.


 

 \newpage

\begin{table}
\centering
\begin{tabular}{ l | c c | c c c c}
   \hline
    & h (nm) & L (nm) & $\Delta\lambda$ (nm)  & $\eta_\text{ext}$ & $F_p$ &FOM \\
  \hline
  low Q & 1375 & 1572  & 14 &  0.3 & 2.2 & 0.57 \\
  medium Q& 1312 & 1502 & 3.3 &  0.5 & 2.4 & 1.3 \\
  high Q & 1200 & 1600 & 0.95 &  0.6 & 6 & 1.85 \\
\end{tabular}
\caption{Cavity parameters, bandwidth and figure of merit for for the three structures considered. The medium-Q configuration features a top DBR with 5 layers and a bottom gold mirror. In the high-Q case, the top DBR consists of 9 layers and the bottom DBR of 20 layers. All DBRs are designed around a central wavelength of 914 nm. In all cases, $R=2L$.}
\label{table:dimensions}
\end{table}

\end{document}